# Hub Star Modeling 2.0
# for Medallion Architecture

**Shahram Salami**[*]
Sydney, Australia
shsalami@gmail.com

**Abstract:** Data warehousing enables performant access to high-quality data integrated from dynamic data sources. The medallion architecture, a standard for data warehousing, addresses these goals by organizing data into bronze, silver and gold layers, representing raw, integrated, and fit-to-purpose data, respectively. In terms of data modeling, bronze layer retains the structure of source data with additional metadata. The gold layer follows established modeling approaches such as star schema, snowflake, and flattened tables. The silver layer, acting as a canonical form, requires a flexible and scalable model to support continuous changes and incremental development. This paper introduces an enhanced Hub Star modeling approach tailored for the medallion architecture, simplifying silver-layer data modeling by generalizing hub and star concepts. This approach has been demonstrated using Databricks and the retail-org sample dataset, with all modeling and transformation scripts available on GitHub.

## 1. Introduction

Enterprise data warehousing presents challenges such as data integration, data quality improvement, change management, and incremental development. Modern data platforms address these challenges through layered architectures (Ravat & Zhao, 2019), recently formalized as the medallion architecture, consisting of bronze, silver, and gold layers (Bhatt & Sekar, 2022).

This architecture employs different modeling paradigms to transform data across layers. The bronze layer ingests raw source data with metadata enrichments. The silver layer consolidates multiple sources into an integrated data model with improved data quality. This model provides a unified view of data (Inmon, Haines, & Rapien, 2022) while accommodating continuous changes and incremental development. Finally, silver-layer data is transformed into structured gold-layer formats, such as star schema, snowflake, and flattened tables. Each gold format requires specific transformations from the silver layer. Additionally, the gold layer enhances change management by eliminating direct access to silver data.

The silver layer, serving as the canonical model, plays a key role in the medallion architecture. However, there is no widely accepted standard for its structure and design. This paper introduces an enhanced Hub Star modeling for the silver layer—a superset of the earlier approach (Salami, 2024). Unlike the previous approach, which applied different structures for various data categories (Van Gils, 2023), this unified model consistently handles reference, master, and transactional data. Its simplified and flexible structure offers a lightweight framework that promotes common terminologies and design patterns in data modeling and development.

Hub Star modeling consists of two main components: hubs and stars. Hubs integrate entities from multiple sources, while stars represent relationships between hubs. Additionally, virtual hubs, including the Time hub and Item hub, support temporal data and weak entities, respectively. Participating Time hub, one star table can capture historical data, whereas hubs store only current attributes.

This paper utilizes Databricks as the data platform and the retail-org, a sample dataset in this platform, to demonstrate the proposed Hub Star modeling. All modeling and transformation scripts are available on GitHub[†]. The paper is structured as follows:
- Section 2 reviews related work.
- Section 3 describes the bronze layer and its raw data.
- Section 4 introduces Hub Star modeling for the silver layer.
- Section 5 discusses the gold layer and presents sample transformations into a star schema.
- Section 6 concludes the paper.

---

[*] https://www.linkedin.com/in/shahram-salami/
https://orcid.org/0000-0003-2200-2042

[†] https://github.com/shsalami/retail-org.git

## 2. Related Work

Hub Star modeling (Salami, 2024) has been introduced for data integration, using reference hubs for reference data and domain hubs for master and transactional data. Star tables represent relationships between hubs, while domain hubs exclude non-key attributes to ensure that data history is maintained in the star tables.

This new version of Hub Star modeling, a superset of the previous version, introduces a generalized hub that includes non-key attributes. In the medallion architecture, data history is preserved in the bronze layer. Since hub tables store only a subset of current data, historical data from the bronze layer can be incrementally transformed into star tables as needed.

## 3. Raw Data in the Bronze Layer

The bronze layer functions as the data ingestion layer, like the staging area in traditional data warehousing. However, unlike a staging area, the bronze layer typically stores ingested data permanently. Additionally, it enriches raw data with metadata attributes, which often include:
- *capture_timestamp*: The timestamp when the change was captured.
- *load_timestamp*: The timestamp when the record was loaded into the bronze table.
- *extract_path*: The source extraction path (e.g., input file path).
- *delete_flag* (*Optional*): Indicate whether the record was deleted.

The capture_timestamp value can be determined based on the following priority order:
- CDC (Change Data Capture) timestamp from the source database
- Last modified timestamp of the record in the source system
- Timestamp when the record was captured in the pipeline

To illustrate data modeling across different layers, this paper uses the retail-org dataset in Databricks as the data source. Figure 1 presents the entity reference diagram of this dataset, where arrows indicate referencing entities pointing to their referenced counterparts. The table sales_orders partially references the customers table, referencing only customer_id, which is part of the composite primary key (customer_id, valid_from) in customers.

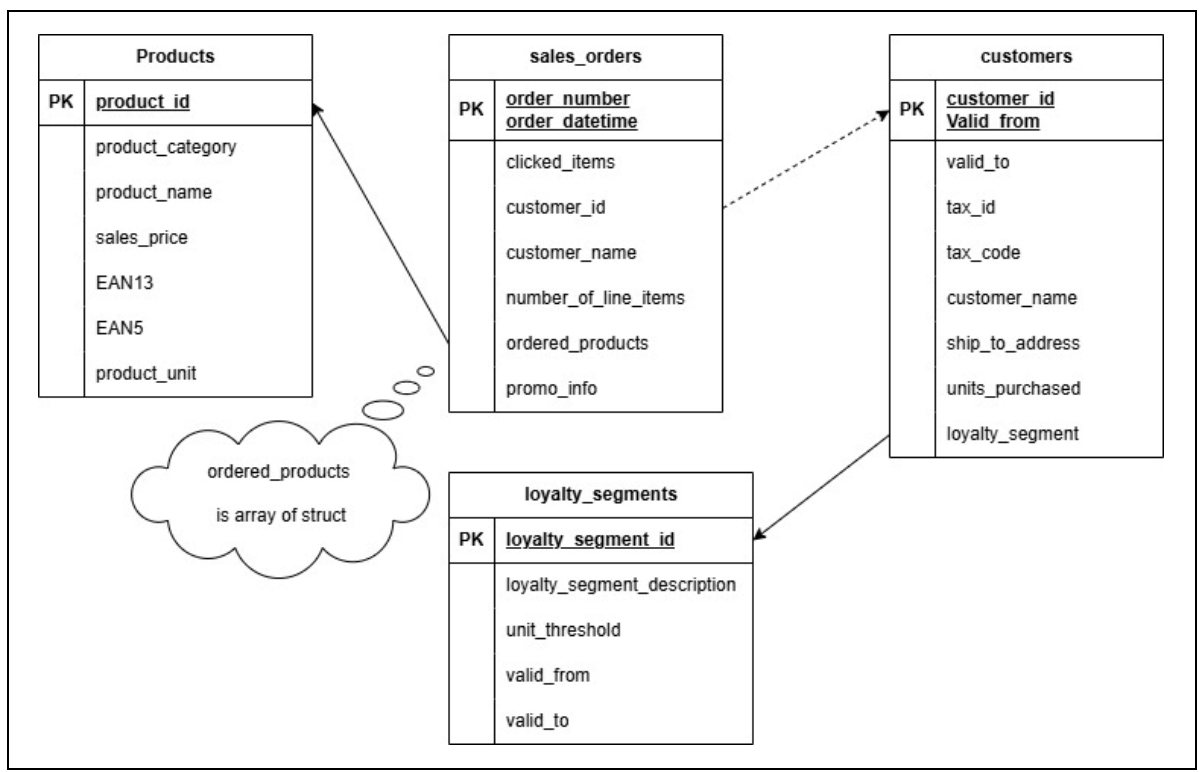

*Figure 1. The entity reference diagram of retail-org dataset in Databricks (/databricks-datasets/retail-org/).*

Streaming tables are a common option for data ingestion in Databricks. The following script demonstrates how to ingest the customers table (in Figure 1) into the bronze layer:

```
CREATE OR REFRESH STREAMING TABLE customers
COMMENT "The customers table, ingested from /databricks-datasets/retail-org."
TBLPRPERTIES ("quality" = "bronze") AS
SELECT _metadata.file_modification_time AS capture_timestamp,
       CURRENT_TIMESTAMP() AS load_timestamp,
       _metadata.file_path AS extract_path,
       *
  FROM CLOUD_FILES("/databricks-datasets/retail-org/customers/", "csv",
                   MAP("cloudFiles.inferColumnTypes", "true"));
```

The script uses the file modification time of the input file as the capture_timestamp, indicating when the record was captured or extracted in the pipeline.

Creating the customers table in Databricks requires specifying the catalog and schema. Distributing objects across different schemas simplifies administration, including change management and security. In the context of medallion architecture, objects can be distributed across bronze, silver, and gold schemas. In contrast, data mesh (Dehghani, 2020) architecture organizes objects into different data products. Naming our sample data product as 'retail', Table 1 outlines schema names used in this paper. In practice, Databricks notebooks can be parameterized to specify the target schemas.

| Layer | Schema | Comment |
|---|---|---|
| Bronze | raw_retail | raw stands for raw data |
| Silver | hs_retail | hs stands for hub star |
| Gold | ss_retail | ss stands for star schema |

*Table 1. Schema names for the sample data product (retail) across different layers.*

## 4. Hub Star Modeling in the Silver Layer

This section formally introduces Hub Star modeling in the silver layer, covering hub and star components.

### 4.1. Hubs

Hubs integrate business entities into the model, each identified by a business key.

**Definition 1:**
*Hub* table defines a business entity and contains:

- *Hub Key*: A simple (non-composite) surrogate key for each row, which can be either system-generated or computed from business keys.
- *Global/ Local Business Key*: One of the immutable and mandatory natural keys that is unique either globally or locally within each data source. Composite business keys consist of multiple attributes.
- *Descriptives (Optional)*: Attributes that store the current version of the data, including the keys of other hubs.
- *Constraints*:
    o The hub key is the table's primary key.
    o A unique constraint is defined for the global business keys.
    o A composite unique constraint is defined on both load_source metadata and local business keys.
    o Business keys must have value.
- *Metadata*:
    o *load_source*: The identifier for the loading source.
    o *capture_timestamp*: The timestamp when the record was captured, with the same value in the bronze table.
    o *load_timestamp*: The timestamp when the record was loaded in the hub (insert/update).
    o *initial_capture_timestamp*: The initial capture timestamp of the record.
    o *delete_flag* (*Optional*): Indicates whether the record is deleted. ▪

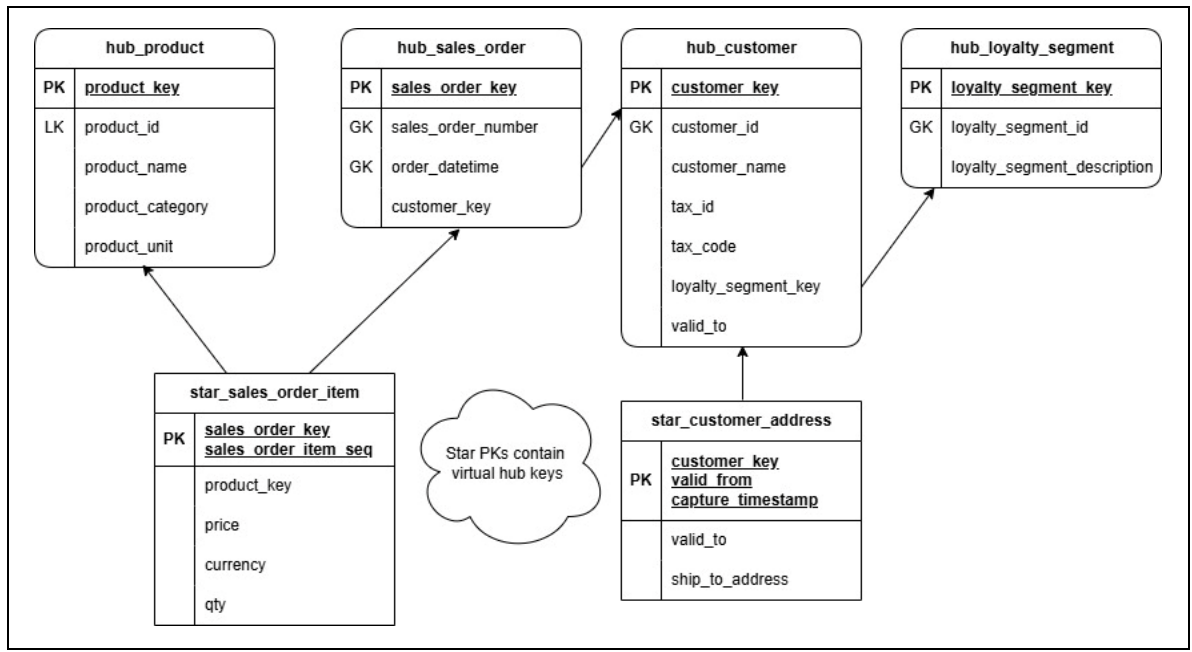

*Figure 2. The Hub Star diagram of the data source in Figure 1. Hubs are presented with curved rectangles. The PK, GK, and LK signs stand for Primary Keys and Global/Local Business Keys, respectively. Metadata are not presented in this diagram.*

Figure 2 represents the Hub Star modelling of the source tables shown in Figure 1. The customer_id in the source customers table serves as the business key attribute. This business key becomes the global business key in the hub_customer table, ensuring uniqueness across all sources. While the customers table contains historical customer data, the hub_customer captures only the current version of the attributes. For example, the customer_name in this hub reflects the most recent name of the customer. Below is the creation script for the hub_customer table in Databricks:

```
CREATE TABLE hs_retail.hub_customer
( load_source INT,
  capture_timestamp TIMESTAMP,
  load_timestamp TIMESTAMP,
  initial_capture_timestamp TIMESTAMP,

  customer_key STRING PRIMARY KEY RELY,
  customer_id INT NOT NULL,
  customer_name STRING,
  tax_id STRING,
  tax_code STRING,
  loyalty_segment_key INT FOREIGN KEY REFERENCES hs_retail.hub_loyalty_segment RELY,
  valid_to TIMESTAMP
);
```

In this script, metadata attributes are placed in the first columns to preserve their positions against future changes.

Constraints such as primary and foreign keys in Databricks are informational and not enforced at the storage level. However, reporting tools can use them to navigate the data model, and the 'RELY' option enables the query optimizer to leverage these constraints for query planning.

**Definition 2:**
*Hub's default rows* are rows in the hubs that use special keys to avoid nulls in the referencing tables. These rows have a default minimum timestamp for the metadata timestamps. ▪

At least one default row with a special hub key exists in each hub to avoid null keys in the referencing tables, which simplifies the model and queries. For example, a null key cannot be part of the primary key in the referencing table. Additionally, the minimum timestamps in default rows can serve as the load high-water mark during the hub's initial load.

As an example, the following script inserts the default row into hub_customer, with the hub key and business key set to -1, and it uses '0' as the system load_source:

```
INSERT INTO hs_retail.hub_customer (
      load_source, capture_timestamp, load_timestamp, initial_capture_timestamp, delete_flag,
      customer_key, customer_id, customer_name, tax_code string, loyalty_segment_key)
VALUES (0, '1970-01-01', '1970-01-01', '1970-01-01', 0, '-1', -1, 'null', 'null', -1);
```

Hub tables can be loaded using the MERGE statement, with capture_timestamp as the load high-water mark. For example, the following script loads hub_customer:

```
MERGE INTO hs_retail.hub_customer hc
USING (SELECT capture_timestamp,
              customer_id,
              customer_name,
              tax_id,
              tax_code,
              loyalty_segment,
              ROW_NUMBER() OVER(PARTITION BY customer_id
                               ORDER BY valid_from DESC, capture_timestamp DESC) rn
        FROM raw_retail.customers
       WHERE capture_timestamp > (SELECT MAX(capture_timestamp) FROM hs_retail.hub_customer)
      ) c
   ON hc.customer_id = c.customer_id
 WHEN MATCHED AND c.rn = 1
             AND (hc.customer_name IS DISTINCT FROM c.customer_name
                  OR hc.tax_id IS DISTINCT FROM c.tax_id
                  OR hc.tax_code IS DISTINCT FROM c.tax_code
                  OR hc.loyalty_segment_key IS DISTINCT FROM c.loyalty_segment
                  OR hc.valid_to IS DISTINCT FROM c.valid_to
                 ) THEN
   UPDATE SET hc.capture_timestamp = c.capture_timestamp,
              hc.load_timestamp = CURRENT_TIMESTAMP(),
              hc.customer_name = c.customer_name,
              hc.tax_id = c.tax_id,
              hc.tax_code = c.tax_code,
              hc.loyalty_segment_key = c.loyalty_segment,
              hc.valid_to = c.valid_to
 WHEN NOT MATCHED THEN
    INSERT (load_source, capture_timestamp, load_timestamp, initial_capture_timestamp,
      customer_key, customer_id, customer_name, tax_id, tax_code, loyalty_segment_key, valid_to)
    VALUES (1, c.capture_timestamp, CURRENT_TIMESTAMP(), c.capture_timestamp,
      SHA2(CAST(customer_id AS STRING), 256),
      c.customer_id, c.customer_name, c.tax_id, c.tax_code, c.loyalty_segment, c.valid_to);
```

**4.2. Hub Key Types**

According to Definition 1, hubs can be identified by either system-generated or computed keys. Table 2 compares these key types.

| **Feature** | **Type1** | **Type2** | **Comment** (for **No** support on the feature) |
|---|---|---|---|
| Always numeric | Yes | No | Non-numeric/composite business keys make char strings |
| Source independent | Yes | No | Affected by source changes on structure of business keys |
| Platform independent | No | Yes | Needs sequence object or auto-increment (identity) columns |
| Same keys on reload | No | Yes | Values depend on the load order of records |
| Lookup-free referencing | No | Yes | Needs hub lookup and key generation for new business keys |
| No hash requirement | Yes | No | Needs hashing to obscure confidential business keys |

*Table 2. Comparison of hub key types: Type1 (System-generated) and Type2 (Computed from business keys)*

In our sample scripts, hubs are defined by computed keys. Table 3 represents the formula used to define different hub keys. In this table, assuming customer_id is private information in customer data, hub_customer applies the SHA2 hash function in Databricks to obscure the customer_key in the referencing tables.

Computed keys typically concatenate parts of composite business keys. For example, the hub key of hub_sales_order is computed using the CONCAT_WS function in Databricks, with # as the delimiter between key parts. By fixing the length of key parts, delimiters can be eliminated, though this is generally impractical.

Similarly, assuming multiple sources for products, hub_product generates a unique hub key by concatenating the local business key with load_source metadata.

Finally, hub_loyalty_segment uses the exact business key value as the hub key. In this case, the hub key acts as a placeholder, simplifying future changes in the hub. For example, loyalty_segment_id may later be changed to a local business key.

| Hub | Business Key (BK) | BK Scope | Hub Key Formula |
|---|---|---|---|
| hub_customer | customer_id | Global | SHA2 (CAST(customer_id AS STRING), 256) |
| hub_sales_order | order_number, order_datetime | Global | CONCAT_WS( '#', order_number, DATE_FORMAT( order_datetime, 'yMMddHHmmss')) |
| hub_product | product_id | Local | CONCAT_WS ('#', load_source, product_id) |
| hub_loyalty_segment | loyalty_segment_id | Global | loyalty_segment_id |

*Table 3. Hub key formula for different hubs in the model*

**4.3. Virtual Hubs**

This section introduces the Time hub and the Item hub as virtual hubs. Unlike other hubs, virtual hubs are not explicitly defined in the model; however, their hub keys are referenced like those of other hubs. These references will be explained further in the next section.

**Definition 3:**
*Time Hub* is a virtual hub that includes all timestamps. ▪

All timestamp attributes (e.g. order_datetime) and metadata timestamps (e.g. capture_timestamp) serve as keys in the Time hub. The precision of timestamps varies based on usage. When certain components are absent, they are assumed to be zero. For example, birth_date typically includes only the date component, meaning its hour, minute, second, and millisecond values are treated as zero.

**Definition 4:**
*Item Hub* is a virtual hub where the hub key corresponds to the sequence of items in a collection, functioning as a weak entity with no independent business key. Each item in the collection is identified using the business keys of other entities along with a partial key representing its sequence within the collection, which serves as the Item Hub key. If no specific sequence attribute exists, the hub key is computed by concatenating (and optionally hashing) a minimal subset of the collection's attributes. ▪

For example, the ordered_products attribute in Figure 1 represents a weak entity, defined as a collection of ordered items within the sales_orders table. The sequence of items ordered is the Item hub key. Without an independent business key for each ordered item, ordered_products remains a weak entity, even if the source system defines it in a separate table rather than as a collection attribute.

**4.4. Stars**

Star tables represent data in the relationships between hubs including virtual hubs.

**Definition 5:**
*Star* table represents data in the relationships between hubs and contains:

- *Hub keys*: Keys of participating hubs, including the virtual hubs.
- *Descriptives (Optional)*: Attributes observed or measured in the relationship.
- *Constraints*: A minimal subset of hub keys (including virtual hubs) forms the composite primary key. This key may include *capture_timestamp* metadata to track *version history* for mutable sources.
- *Metadata*:
    - *load_source*: Identifier of the loading source.
    - *capture_timestamp*: The timestamp when the record was captured, with the same value in the bronze table.
    - *load_timestamp*: The timestamp when the record was inserted in the star.
    - *delete_flag* (*Optional*): Indicates whether the record is deleted. ▪

Restricting the star composite key to a subset of hub keys improves model readability by clearly defining star relationships between hubs and promoting better design by considering the necessary hubs.

In Figure 1, the customers table represents the history of customer attributes using the valid_from and valid_to columns. For simplicity, address attributes other than ship_to_address are not shown in the figure. The customer address history is modeled by star_customer_address in Figure 2. The composite primary key of this table includes the valid_from attribute and capture_timestamp metadata (as instances of a virtual Time hub) to represent the history of customer addresses.

The star_sales_order_item table is another star table in Figure 2, modeling the ordered_products array from the sales_order table in Figure 1. The sales_order_item_seq represents the sequence of each ordered item within the array and is added to the primary key of the star table, acting as a virtual item hub. Using LATERAL VIEW EXPLODE to access array items and the TRANSFORM lambda function to determine the sequence of each item, this star table can be implemented as a streaming table in Databricks:

```
CREATE OR REFRESH STREAMING TABLE star_sales_order_item
( load_source INT,
  capture_timestamp TIMESTAMP,
  load_timestamp TIMESTAMP,

  sales_order_key STRING NOT NULL FOREIGN KEY REFERENCES hs_retail.hub_sales_order RELY,
  sales_order_item_seq INT NOT NULL,
  order_datetime TIMESTAMP,
  product_key STRING FOREIGN KEY REFERENCES hs_retail.hub_product RELY,
  price BIGINT,
  currency STRING,
  qty INT,
  CONSTRAINT star_sales_order_item_pk PRIMARY KEY (sales_order_key, sales_order_item_seq) RELY
AS
SELECT 1 load_source, so.capture_timestamp, CURRENT_TIMESTAMP() load_timestamp,
       CONCAT_WS('#', so.order_number, DATE_FORMAT( TIMESTAMP( FROM_UNIXTIME(
                CAST(so.order_datetime AS LONG))), 'yMMddHHmmss')) sales_order_key,
       op.item_seq as sales_order_item_seq,
       TIMESTAMP(FROM_UNIXTIME(CAST(so.order_datetime AS LONG))) order_datetime,
       CONCAT_WS('#', 1, op.item.id) product_key,
       op.item.price, op.item.curr currency, CAST(op.item.qty AS INT) qty
  FROM STREAM(raw_retail.sales_orders) AS so LATERAL VIEW
       EXPLODE(TRANSFORM(ordered_products,(x,i) -> STRUCT(x AS item,(i+1) AS item_seq))) AS op;
```

Generally, stars with the same participating hubs and composite keys can be modelled as a single star to avoid extra joins. However, separate stars may be preferred in certain cases to:

- Reduce data volume by separating attributes with different update rates.
- Simplify data loading by separating attributes from different sources.
- Minimize null columns by separating some nullable attributes.
- Comply with security agreements by isolating confidential attributes.
- Address platform limitations on the number of columns by splitting extra attributes.

For example, star_customer_name can be added to the silver layer to track changes in customer names. Alternatively, both customer_name and ship_to_address attributes can be modeled in a single star table as star_customer_info.

### 4.5. History Options

Star tables can represent version history and life cycle of records, while hubs store the current version of attributes. Generally, a source attribute can be modeled in a hub, a star, or both. Table 4 compares the use cases for each scenario.

| Maintaining | Modelled as | Use Case |
|---|---|---|
| Only the current version | Hub Attribute | Immutable attributes |
| | | No analytic requirement for the attribute history |
| | | Partial modeling of new entities |
| Version history | Star Attribute | Versioning of mutable attributes |
| | | Representing reactivation history of deleted records |
| | | Handling multivalued attributes (e.g. customer phone numbers) |
| | Hub + Star | Caching the current version of an attribute in the hub |
| | | Supporting backward compatibility for existing hub attributes |

*Table 4. Use cases for different types of attribute modeling*

### 4.5.1. Maintaining Only the Current Version

If there is no analytic requirement to track the history of an attribute, it can be modeled as a hub attribute, representing the current version of the data. Table 5 shows the bronze-to-silver entity mapping for the sample model. Typically, immutable reference data attributes are stored in hub tables, such as hub_loyalty_segment. Additionally, assuming there is no business case for tracking the history of product attributes, they are modeled in hub_product.

| Bronze Table | Hub Table | Star Table |
|---|---|---|
| loyalty_segemnts | hub_loyalty_segments | - |
| products | hub_product | - |
| customers | hub_customer | star_customer_address |
| sales_orders | hub_sales_order | star_sales_order_item |

*Table 5. Bronze (Figure 1) to silver (Figure 2) entity mappings.*

Even when there is no business use case for historical data, the bronze layer typically retains the data history. This historical data can be leveraged for change tracking and auditing, and it also supports the incremental development of historical data in star tables.

### 4.5.2. Maintaining Version History

The star table can represent version history of attributes by including capture_timestamp in the star composite key. Some transactional data are in append-only mode with no requirement for historical versioning. For example, star_sales_order_item does not support historical versioning, but the entire set of ordered items can be expired by deleting the corresponding sales_order_key in the hub_sales_order.

Reactivation history of records is a special case of version history in star tables. While hubs are not designed to track history of records, their primary purpose is to map business keys to hub keys. For instance, once a customer address is flagged as deleted in star_customer_address, it can later be reactivated by inserting a new record with the same values. In contrast, reactivating a deleted product in hub_product involves updating the existing record. However, delete operation is uncommon for reference and master data entities, which often use special attributes to indicate logical deactivation, such as valid_to in hub_customer.

Representing an attribute in both hub and star tables can enhance the performance of non-historical queries and ensure backward compatibility when accessing hub attributes. However, if there are no performance concerns, it is advisable to remove the attribute from the hub table in future development cycles once it has been added to the star table to avoid the extra cost of maintaining both. For example, if customer_name is represented in star_customer_name to track all versions, it becomes a candidate for removal from hub_customer.

## 5. Business Forms in the Gold Layer

BI applications typically use star schema and snowflake models, which consist of fact and dimension tables. Dimensions are categorized into different types of Slowly Changing Dimensions (SCD) (Kimball & Ross, 2013). SCD Type 1 maintains only current records, whereas SCD Type 2 retains both current and historical records.

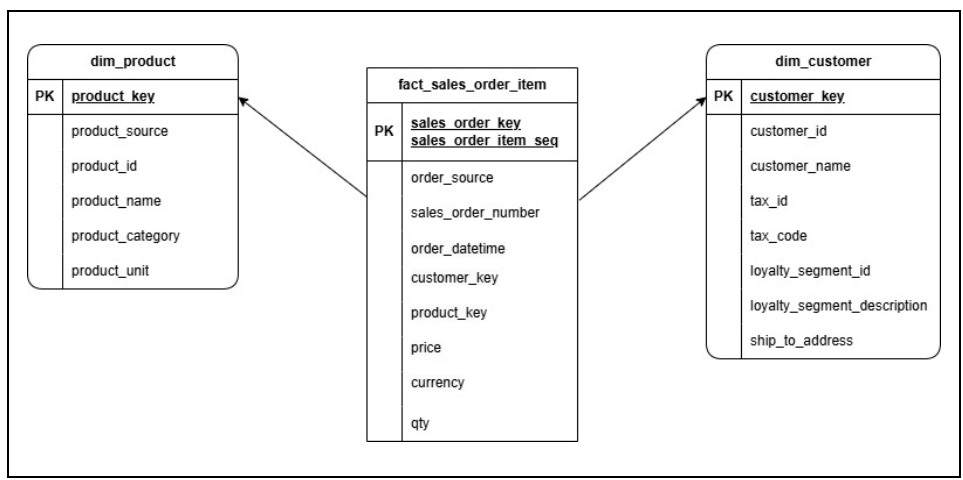

*Figure 3. The FactOrderItem and its SCD type-1 dimensions.*

This section presents simple transformations from the hub star model to fact and dimension tables, as illustrated in Figure 3. Depending on query complexity, these transformations are implemented as either views or materialized views. In Databricks, both views and materialized views offer efficient transformations for the gold layer. Views do not require physical storage or refresh time, while materialized views provide better query performance for performance-critical facts and dimensions.

### 5.1. Transformation to SCD Type 1

As an example of an SCD Type 1 transformation, the following script creates dim_product as a simple wrapper view on hub_product:

```
CREATE VIEW ss_retail.dim_product AS
SELECT product_key,
       load_source product_source,
       product_id,
       product_name,
       product_category,
       product_unit
  FROM hs_retail.hub_product;
```

The dim_customer dimension joins hub_customer with hub_loyalty_segment and the current rows of star_customer_address in the 'sca' query. The following script uses a materialized view to improve performance for this more complex query:

```
CREATE MATERIALIZED VIEW dim_customer AS
WITH sca AS (
SELECT *
  FROM (SELECT *, ROW_NUMBER() OVER(PARTITION BY customer_key
                                    ORDER BY valid_from DESC, capture_timestamp DESC) AS rn
          FROM hs_retail.star_customer_address
       )
 WHERE rn = 1
   AND delete_flag = 0
)
SELECT hc.customer_key,
       hc.customer_id,
       hc.customer_name,
       hc.tax_id,
       hc.tax_code,
       hls.loyalty_segment_id,
       hls.loyalty_segment_description,
       sca.ship_to_address
  FROM hs_retail.hub_customer hc
  JOIN hs_retail.hub_loyalty_segment hls ON hc.loyalty_segment_key = hls.loyalty_segment_key
  LEFT JOIN sca ON hc.customer_key = sca.customer_key;
```

The fact_sales_order_item fact table joins star_sales_order_item with hub_sales_order, which contains the keys for dim_product and dim_customer:

```
CREATE VIEW ss_retail.fact_order_item AS
SELECT soi.sales_order_key,
       soi.sales_order_item_seq,
       hso.sales_order_number,
       hso.order_datetime,
       hso.customer_key,
       soi.product_key,
       soi.price,
       soi.quantity,
       soi.qty
  FROM hs_retail.star_order_item soi
  JOIN hs_retail.hub_sales_order hso ON soi.sales_order_key = hso.sales_order_key;
```

### 5.2. Transformation to SCD Type 2

SCD Type 2 dimensions track data history using capture_timestamp metadata or other effective/valid source timestamps. The following script includes all customer addresses in the dim_customer2 dimension. The dimension key is composed of customer_key and valid_from, which are concatenated to create a non-composite key:

```sql
CREATE MATERIALIZED VIEW dim_customer2 AS
WITH sca AS (
SELECT *
  FROM (SELECT *, ROW_NUMBER() OVER( PARTITION BY customer_key, valid_from
                                     ORDER BY capture_timestamp DESC) AS rn
          FROM hs_retail.star_customer_address
       )
 WHERE rn = 1
   AND delete_flag = 0
)
SELECT CONCAT_WS('#', hc.customer_key, sca.valid_from) customer2_key,
       hc.customer_id,
       hc.customer_name,
       hc.tax_id,
       hc.tax_code,
       hls.loyalty_segment_id,
       hls.loyalty_segment_description,
       sca.ship_to_address,
       sca.valid_from,
       sca.valid_to
  FROM hs_retail.hub_customer hc
  LEFT JOIN hs_retail.hub_loyalty_segment hls
         ON hc.loyalty_segment_key = hls.loyalty_segment_key
  LEFT JOIN sca ON hc.customer_key = sca.customer_key;
```

The new version of fact_order_item also includes the dim_customer2 key:

```sql
CREATE VIEW ss.retail.fact_order_item AS
SELECT soi.sales_order_key,
       soi.sales_order_item_seq,
       hso.sales_order_number,
       soi.order_datetime,
       ho.customer_key,
       dc.customer2_key,
       soi.product_key,
       soi.price,
       soi.quantity,
       soi.qty
  FROM hs_retail.star_sales_order_item soi
  JOIN hs_retail.hub_sales_order hso ON soi.sales_order_key = hso.sales_order_key
  LEFT JOIN ss.retail.dim_customer2 dc
         ON hso.order_datetime BETWEEN dc.valid_from AND dc.valid_to;
```

## 6. Conclusion

This paper introduced hub star modeling within the medallion architecture. Considering the bronze, silver, and gold layers, the updated hub star modeling provides a lightweight framework with design terminologies and guidelines for modeling the silver layer. The proposed approach consists of hub and star components: hubs represent enterprise entities, while stars define data in hub's relationships. Hubs store the current version of attributes, whereas stars preserve data history using Time hub as a virtual hub. Additionally, this paper outlined core metadata attributes for the bronze and silver layers, enhancing data organization and historical tracking within the architecture.

This paper primarily focused on the logical design of the silver layer. The optimization of physical design and transformations between layers depends on the chosen data platform, which will be explored in future work.

# References

[Bhatt & Sekar, 2022] Bhatt, S., & Sekar, D., *Data Warehousing Modeling Techniques and Their Implementation on the Databricks Lakehouse Platform,* (2022). https://www.databricks.com/blog/2022/06/24/data-warehousing-modeling-techniques-and-their-implementation-on-the-databricks-lakehouse-platform.html.

[Dehghani, 2020] Dehghani, Z., *Data mesh priniples and logical architeture*. (2020). https://martinfowler.com/articles/data-mesh-principles.html

[Inmon, Haines, & Rapien, 2022] Inmon, B., Haines, P., & Rapien, D., *Integrating Data,* (2022). Technics Publications.

[Kimball & Ross, 2013] Kimball, R., & Ross, M., *The Data Warehouse Toolkit: The Definitive Guide to Dimensional Modeling 3rd Edition,* (2013). Wiley.

[Ravat & Zhao, 2019] Ravat, F., & Zhao, Y., *Data Lakes: Trends and Perspectives.* International Conference on Database and Expert Systems Applications (pp. 304–314), (2019). https://doi.org/10.1007/978-3-030-27615-7.

[Salami, 2024] Salami S., *Hub Star Modeling for Data Integration*. ResearchGate, (2024). https://www.researchgate.net/publication/380910376_Hub_Star_Modelling_for_Data_Integration/

[Van Gils, 2023] Van Gils, B., *Reference and Master Data Management*. In Data in Context (pp. 135-140), (2023). Springer.